\documentclass[aps,prb,twocolumn,groupedaddress,floatfix,amsmath,amssymb,amsfonts]{revtex4}

\usepackage{graphicx}

\begin{document}

\title{Quantum phase transitions in attractive extended Bose-Hubbard Model with three-body constraint}

\author{Yung-Chung Chen}
\affiliation{Department of Physics, Tunghai University, Taichung
40704, Taiwan}

\author{Kwai-Kong Ng}
\affiliation{Department of Physics, Tunghai University, Taichung
40704, Taiwan}

\author{Min-Fong Yang}
\affiliation{Department of Physics, Tunghai University, Taichung
40704, Taiwan}

\date{\today}

\begin{abstract}
The effect of nearest-neighbor repulsion on the ground-state phase
diagrams of three-body constrained attractive Bose lattice gases
is explored numerically.
When the repulsion is turned on, in addition to the uniform Mott
insulating state and two superfluid phases (the atomic and the
dimer superfluids), a dimer checkerboard solid state appears at
unit filling, where \emph{boson pairs} form a solid with
checkerboard structure.
We find also that the first-order transitions between the uniform
Mott insulating state and the atomic superfluid state can be
turned into the continuous ones as the repulsion is increased.
Moreover, the stability regions of the dimer superfluid phase can
be extended to modest values of the hopping parameter by tuning
the strength of the repulsion. Our conclusions hence shed light on
the search of the dimer superfluid phase in real ultracold Bose
gases in optical lattices.
\end{abstract}


\maketitle


Spectacular progress both in theories and experiments has recently
been made on ultracold atomic and molecular gases in optical
lattices. Owing to the remarkable control over physical
parameters, ultracold gases offer opportunities to simulate the
physics of strongly correlated systems in regimes which are not
easily accessible to solid-state materials. As a result, they
provide very clean and tunable systems in the search for exotic
quantum phases and in probing quantum critical behaviors around
these phases.~\cite{BEC_reviews} For instance, successful
experimental realization of the superfluid-Mott transition for
ultracold bosons in an optical lattice~\cite{BEC_exps} has paved
the way for studying other strongly correlated phases in various
lattice models.

Very recently, it was suggested that intriguing quantum critical
behaviors could occur in attractive bosonic lattice gases with
three-body on-site constraint.~\cite{Diehl10,Lee10} The on-site
constraint can arise naturally due to large three-body loss
processes,~\cite{Daley09,Roncaglia10} and it stabilizes the
attractive bosonic systems against collapse. Such three-body
constrained systems can be realized also in Mott insulating states
of ultracold spin-one atoms at unit filling.~\cite{Mazza10} As
found by the authors in Ref.~\onlinecite{Diehl10}, a dimer
superfluid (DSF) phase consisting of the condensation of boson
pairs can be realized under sufficiently strong attraction.
According to their analysis, the transitions between the DSF phase
and the conventional atomic superfluid (ASF) state are proposed to
be of Ising-like at unit filling and driven first-order by
fluctuations via the Coleman-Weinberg
mechanism~\cite{Coleman-Weinberg} at other fractional fillings.
Later investigation focuses on the nature of the
superfluid-insulator transitions.~\cite{Lee10} It is shown that,
while the Mott-insulator (MI) to DSF transitions are always of
second order, the continuous MI-ASF transitions can be preempted
by first-order ones and interesting tricritical points can thus
appear on the MI-ASF phase boundaries. The conclusions obtained in
Refs.~\onlinecite{Diehl10} and \onlinecite{Lee10} are partly
supported by a recent numerical study employing stochastic series
expansion (SSE) quantum Monte Carlo (QMC) method implemented with
a generalized directed loop algorithm.~\cite{Bonnes11} In
particular, the existence of a tricritical point along the
saturation transition line is verified. However, the nature of
MI-DSF transitions is not examined in their QMC work.

\begin{figure}[tb]
\includegraphics[clip,width=\columnwidth]{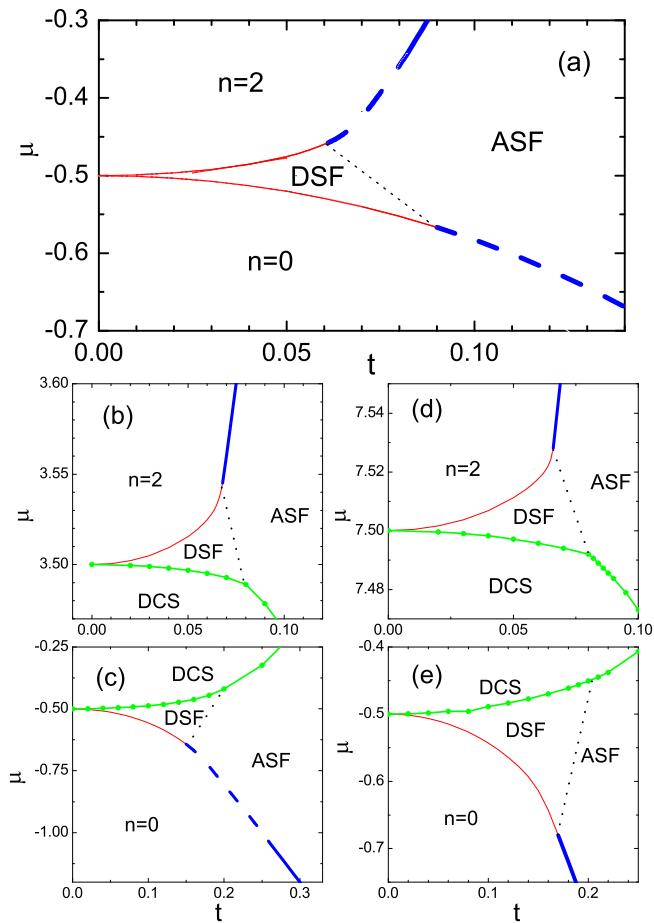}
\caption{(Color online) %
Ground-state phase diagram obtained by ED with (a) $V=0$, (b)(c)
$V=1/2$, and (d)(e) $V=1$. %
The thin red solid lines indicate the continuous MI-DSF
transitions, and the thick blue solid (dashed) lines show the
continuous (first-order) MI-ASF transitions.
The phase boundaries of the DCS state are denoted by the green
solid circles (lines are guides to the eye).
The error bar of the determined phase boundaries is smaller than
the width of the lines or the size of the symbols.
The schematic phase boundaries between the ASF and the DSF phases
are added as the thin dotted lines for clarity. }
\label{fig:phase_diag}
\end{figure}

In the present work, the effect of the nearest-neighbor mutual
repulsion on the ground-state phase diagrams of three-body
constrained attractive lattice bosons is investigated by means of
exact diagonalizations (ED).
The systems under consideration are described by the extended
Bose-Hubbard model with a three-body constraint
$a_i^{\dag\,3}\equiv 0$ on square lattices,
\begin{align}
H &= H_{\rm EBH} - \mu \sum_{i} n_{i} \; , \\
H_{\rm EBH} &= - t \sum_{\langle i,j\rangle} a_{i}^{\dagger }a_{j} %
     + \frac{U}{2} \sum_{i} n_{i}(n_{i} -1)   %
     + V \sum_{\langle i,j \rangle}  n_i n_j \;  . \nonumber
\label{eqn:H}
\end{align}
Here, $a_i (a_i^\dag)$ is the bosonic annihilation (creation)
operator at site $i$, $t$ is the nearest-neighbor hopping
integral, $U<0$ is the on-site two-body attraction ($|U|\equiv 1$
as the energy unit), and $\mu$ is the chemical potential. $V>0$
denotes the nearest-neighbor repulsion, which can come from the
dipole-dipole interactions of the dipolar bosons polarized
perpendicularly to the lattice plane by truncating it at the
nearest-neighbor distance.
For the discussions of possible experimental realizations on the
present model, we refer to Ref.~\onlinecite{Dalmonte11} and
references therein.
The averaged particle density is denoted by $n$ and periodic
boundary conditions are assumed.
Our main results are summarized in Fig.~\ref{fig:phase_diag}. When
$V\neq0$, in addition to the MI and two superfluid phases
mentioned above, a dimer checkerboard solid (DCS) state consisting
of the checkerboard arrangement of \emph{boson pairs} emerges and
occupies the middle part of the phase diagram in the small-$t$
limit. The DSF states now appear only in between the uniform MI
and the DCS states.
Moreover, finite repulsion has interesting effects on the
MI-superfluid transitions also. Because a modest nearest-neighbor
repulsion can avoid cluster formation and then will suppress phase
separation, we observe that the segments of the first-order MI-ASF
transitions on phase boundaries of either the $n=0$ or the $n=2$
MI states (denoted by thick blue dashed lines in
Fig.~\ref{fig:phase_diag}) can shrink to zero as $V$ is increased.
Same conclusion has been reached in other
context.~\cite{Schmidt06} Besides, our findings show that the
phase boundaries of the MI-DSF transitions (denoted by thin red
solid lines in Fig.~\ref{fig:phase_diag}) and therefore the
stability region of the DSF phase in the low-density limit can be
extended to modest hopping parameters $t$ upon tuning $V$, such
that experimental exploration of this interesting state becomes
more feasible.
Our work hence provides a useful guide to the experimental search
of the DSF phase and the associated quantum phase transitions in
ultracold Bose gases in optical lattices.

The details of our analysis are explained below. According to the
discussions in Ref.~\onlinecite{Schmidt06}, one can determine the
phase boundaries of either the $n=0$ MI state (i.e., the empty
state) or the $n=2$ MI state (i.e., the completely filled state)
for a given $t$ as follows. For the transitions out of the $n=0$
MI state, we calculate the excitation energies (relative to the
$n=0$ state) $E(1)-\mu$, $E(2)-2\mu$, and $E(4)-4\mu$ of the
lowest states within the subspaces of fixed total particle numbers
$N_{\rm p}=1$, 2, and 4, respectively. Here $E(N_{\rm p})$ denotes
the lowest excitation energy of $H_{\rm EBH}$ within the fixed
$N_{\rm p}$ subspace. From these quantities, the two- and the
four-particle binding energies, $\Delta_{\rm 2p}=E(2)-2E(1)$ and
$\Delta_{\rm 4p}=E(4)-2E(2)$, can be obtained. For a given value
of the hopping parameter $t$, if both $\Delta_{\rm 2p}$ and
$\Delta_{\rm 4p}$ are positive, there exists no bound states and
the energy gap of the single-particle state closes first upon
increasing the chemical potential $\mu$. This leads to a
continuous MI-ASF transition at $\mu_{c,\textrm{0MI-ASF}} =
E(1)$.~\cite{note1} When $\Delta_{\rm 2p}<0$ but $\Delta_{\rm
4p}>0$, instead, a bound state of two particles appears and its
gap closes first. Thus a continuous MI-DSF transition will occur
at $\mu_{c,\textrm{0MI-DSF}} = E(2)/2$. Aside from these two
possibilities, the condition of $\Delta_{\rm 4p}<0$ gives a
precursor of instability of boson pairs towards cluster formation.
That is, phase separation emerges and a first-order transition
will be observed in varying $\mu$. In the present ED analysis, we
estimate the first-order transition point by $E(4)/4$, which
provides an upper bound of the exact value. Similar discussions
apply to the transitions out of the $n=2$ MI state, where the
holes created from the $n=2$ state take the role played by the
particles discussed previously. We remind that the first-order
transition points estimated by the four-hole excitation energies
will be instead lower bounds of the exact values along the
saturation transition line.

\begin{figure}[tb]
\includegraphics[clip,width=\columnwidth]{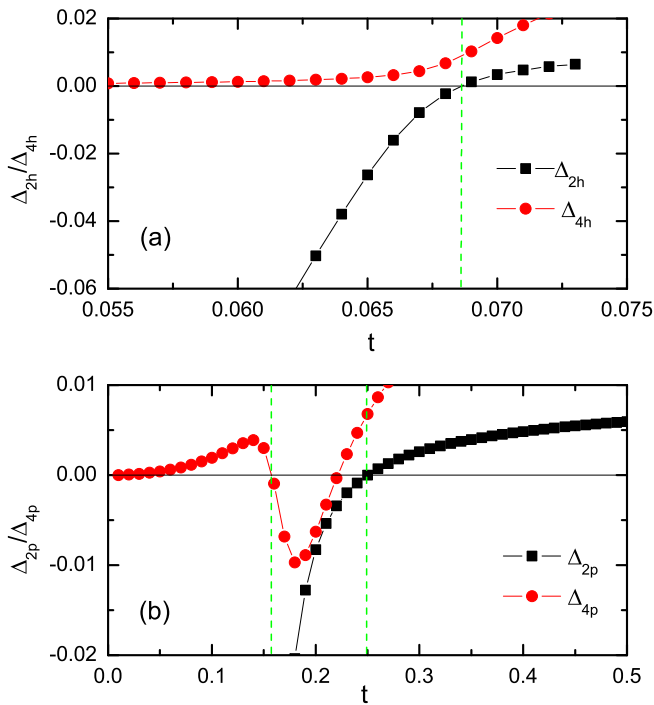}
\caption{(Color online) %
(a) Two- and four-hole binding energies, $\Delta_{\rm 2h}$ and
$\Delta_{\rm 4h}$, around the $n=2$ state and (b) two- and
four-particle binding energies, $\Delta_{\rm 2p}$ and $\Delta_{\rm
4p}$, around the $n=0$ state as functions of $t$ for $V=1/2$ and
$N_{\rm s}=10\times10$. The dashed lines separate the transitions
of different characters. } \label{fig:binding_energies}
\end{figure}

Here distinct MI-superfluid transitions are determined by the
above method for systems of $N_{\rm s}=10\times10$ sites. The
analysis for the $V=1/2$ case is described below for illustration.
The results of various binding energies as functions of hopping
parameter $t$ for this $V$ are presented in
Fig.~\ref{fig:binding_energies}. As seen from
Fig.~\ref{fig:binding_energies}(b), there is a finite region of
$t$ within which $\Delta_{\rm 4p}<0$. It implies that, on the
phase boundary of the $n=0$ state, the continuous MI-DSF (for
$\Delta_{\rm 2p}<0$ but $\Delta_{\rm 4p}>0$) and continuous MI-ASF
(for both $\Delta_{\rm 2p}$, $\Delta_{\rm 4p}>0$) transitions are
separated by first-order MI-ASF transitions for $0.15 \lesssim t
\lesssim 0.26$. On the contrary, the two continuous MI-superfluid
transitions on the phase boundary of the $n=2$ state should meet
directly at $t\approx 0.068$. The transition points $\mu_c$ for a
given $t$ can be determined by using the excitation energies
$E(N_{\rm p})$ as explained above.~\cite{note2}
According to our ED calculations, for $V=1/2$, the continuous
$n=2$ MI-DSF transitions end at
$(t_{\rm E}, \mu_{\rm E}) \simeq (0.068, 3.55)$, %
while $(t_{\rm E}, \mu_{\rm E}) \simeq (0.15, -0.64)$ %
for that of the continuous $n=0$ MI-DSF transition line. There
exists also a tricritical point %
$(t_{\rm T}, \mu_{\rm T}) \simeq (0.26, -1.04)$ %
on the $n=0$ MI-ASF transition line separating the continuous from
the first-order ones.

\begin{figure}[tb]
\includegraphics[clip,width=\columnwidth]{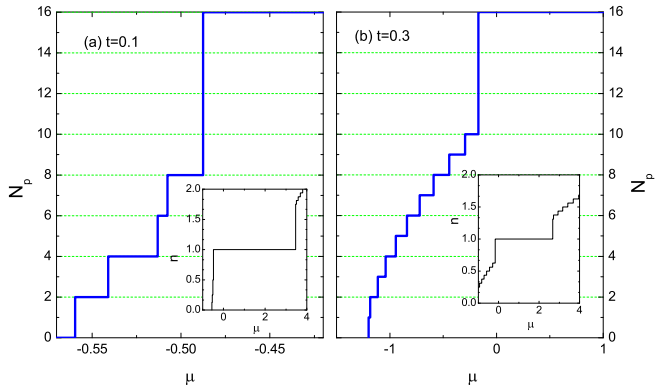}
\caption{(Color online) %
Total particle numbers $N_{\rm p}$ as functions of chemical
potential $\mu$ for (a) $t=0.1$ and (b) $t=0.3$ with $V=1/2$ and
system sizes $N_{\rm s}=4\times4$.
The insets illustrates the complete changes in particle density
$n$ from the $n=0$ to the $n=2$ states as $\mu$ increases. }
\label{fig:n_rho_vs_mu}
\end{figure}


We now turn to the discussions on the phase boundaries of the DCS
state. Due to the effect of the nearest-neighbor repulsion $V$,
\emph{boson pairs} can form a solid with checkerboard structure at
unit filling $n=N_{\rm p}/N_{\rm s}=1$ and lead to the DCS state.
In the zero-hopping limit, this DCS state can be stabilized when
$-1/2 < \mu < -1/2 + 2zV$ with the coordination number $z=4$ for
square lattices. The DCS state can melt into the DSF or the ASF
states under increasing hopping, and its stability region in $\mu$
is expected to reduce to zero as $t$ increases. This picture is
supported by our numerical calculations as seen in
Fig.~\ref{fig:phase_diag}. Due to the limitation in numerics, the
phase boundaries of the DCS state are estimated from the $n=1$
plateaus in the $\mu$-$n$ plots for different $t$'s with systems
sizes of $N_{\rm s}=4\times4$. For illustration, total particle
numbers $N_{\rm p}$ and the particle density $n$ as functions of
chemical potential $\mu$ for two different values of $t$'s with
$V=1/2$ are presented in Fig.~\ref{fig:n_rho_vs_mu}. The large
plateaus at $n=1$ in the $\mu$-$n$ plots clearly indicate the
presence of the DCS state. Two ends of the plateau give the
melting transition points in $\mu$ for a given $t$, as depicted in
Fig.~\ref{fig:phase_diag}.
Moreover, distinct characters in various transitions can be
revealed by the way in which the total particle number $N_{\rm p}$
changes upon varying the chemical potential
$\mu$.~\cite{Schmidt06} In the conventional ASF phase, the number
of particles will increase by 1 when the chemical potential is
increased. However, in the DSF phase, only pairs of bosons appear
in the system due to the presence of a pairing gap. Thus adding a
single boson is forbidden and the jumps in the particle number by
2 will be observed as $\mu$ is varied. Besides forming pairs,
bosons can become unstable towards cluster formation such that the
particle number jumps by a finite amount in the course of tuning
$\mu$. This corresponds to a first-order transition under the
change in the chemical potential. As shown in
Fig.~\ref{fig:n_rho_vs_mu}, for $t=0.1$, the system evolves across
a continuous MI-DSF transition from the $n=0$ state to the DSF
state, and then follows a first-order transition to the DCS state.
For larger $t$ (say, $t=0.3$), the state after the continuous
transition from the $n=0$ MI state becomes the ASF one instead.
Adding more bosons by further increasing $\mu$, the system can
again follows a first-order transition to the DCS state.

\begin{figure}[tb]
\includegraphics[clip,width=\columnwidth]{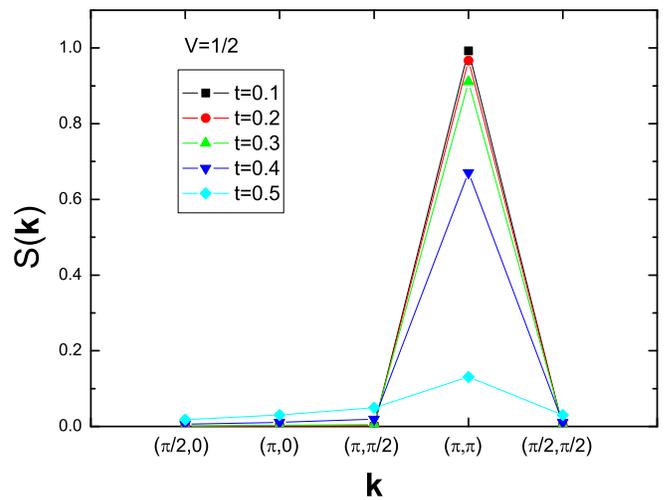}
\caption{(Color online) %
Static structure factor $S(\mathbf{k})$ of the $n=1$ states for
various $t$'s with $V=1/2$ and $N_{\rm s}=4\times4$. }
\label{fig:structure_factor}
\end{figure}

To provide support on the nature of the DCS states within the
$n=1$ plateaus in the $\mu$-$n$ curves, the static structure
factors $S(\mathbf{k})=(1/N_{\rm s}^2) \langle |\sum_j n_j
e^{i\mathbf{k}\cdot\mathbf{r}_j}|^2 \rangle$ of the $n=1$ states
for several hopping parameters $t$'s with $V=1/2$ and system size
$N_{\rm s}=4\times4$ are shown in Fig.~\ref{fig:structure_factor}.
It is found that the static structure factors do have peaks at the
wave vector $\mathbf{k}=(\pi,\pi)$ and thus signal the
checkerboard pattern of the boson pairs. As observed from
Fig.~\ref{fig:structure_factor}, when $t$ is increased, the peak
value $S(\pi,\pi)$ of the structure factor will decrease from its
classical value $S(\pi,\pi)=1$ at $t=0$. This indicates the
quantum melting of the DCS state into the DSF or the ASF states
under increasing hopping.

\begin{figure}[t]
\includegraphics[clip,width=\columnwidth]{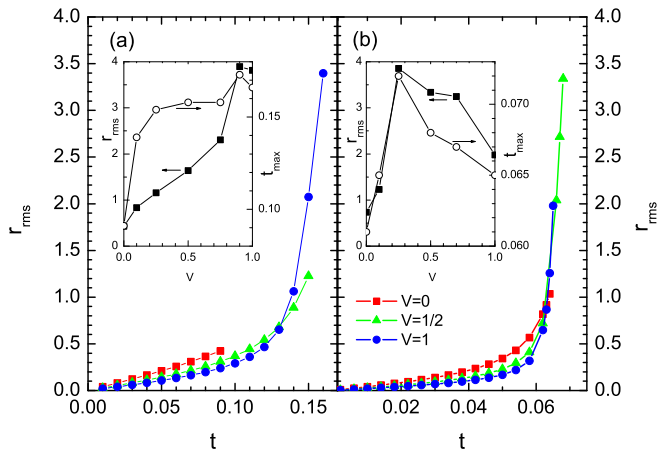}
\caption{(Color online) %
Root-mean-square separation $r_{\rm rms}$ of the (a) particle (b)
hole pairs on $N_{\rm s}=10\times10$ square lattices as functions
of $t$ for various repulsions $V$. %
Insets: $r_{\rm rms}$ and maximal extent $t_{\rm max}$ in hopping
integral for the MI-DSF transitions as functions of $V$.}
\label{fig:coherence_length}
\end{figure}


Some further microscopic details can be uncovered by the ED
calculations. Similar to the counterpart of fermion pairing, a
smooth crossover from the on-site pairs to the loosely bound pairs
may occur also in the present attractive boson systems as the
hopping parameter $t$ increases. In
Fig.~\ref{fig:coherence_length}, evidences supporting this
expectation are presented. Here the coherence length of the boson
pairs is estimated by the root-mean-square separation $r_{\rm
rms}\equiv \sqrt{\langle r^2 \rangle}$,~\cite{Ohta1995,Leung02}
which is evaluated under the ground state of a single pair within
the two-particle or the two-hole subspaces. Our results show rapid
but smooth crossovers from the tightly bound molecules to the
loosely bound pairs, as long as the DSF phase can be stabilized up
to modest values of $t$ (say, the $V=1$ case for the particle
pairs and the $V=1/2$ case for the hole pairs). The general
dependence on the repulsion $V$ of the maximal extent $t_{\rm
max}$ in hopping integral for the stable DSF state in the
low-density limit and the corresponding $r_{\rm rms}$ is shown in
the insets of Fig.~\ref{fig:coherence_length}.
We find that larger $t_{\rm max}$'s in general lead to longer
$r_{\rm rms}$'s. The dependence of $t_{\rm max}$ and $r_{\rm rms}$
on $V$ is found to be nonmonotonic, and their functional forms
shows asymmetry between the cases of the particle and the hole
pairs.
Our conclusions presented in Figs.~\ref{fig:phase_diag} and
\ref{fig:coherence_length} should be of help in determining
optimal experimental settings in the search of the DSF phase in
ultracold Bose gases in optical lattices.


In summary, the ground-state phase diagrams of the three-body
constrained extended Bose-Hubbard model for various repulsions are
investigated. Large plateaus at $n=1$ in the $\mu$-$n$ curves
which show the DCS states are observed. Finite repulsions modify
the MI-superfluid transitions also. We find that the repulsion $V$
can change the first-order MI-ASF transitions into the continuous
ones and the stability regions of the DSF phase can be tuned by
$V$. Therefore, carefully adjusting system parameters into the
suggested parameter regime are necessary to find experimentally
the interesting DSF phase in real ultracold Bose gas in optical
lattices.

Y.-C.C., K.-K.N., and M.-F.Y. thank the National Science Council
of Taiwan for support under Grant No. NSC 99-2112-M-029-002-MY3,
NSC 97-2112-M-029-003-MY3, and NSC 99-2112-M-029-003-MY3,
respectively.


\begin{thebibliography}{32}
\bibitem{note2}
For the $V=0$ case, the phase boundaries of the continuous
transitions achieved by ED coincide with the results evaluated by
the SSE QMC algorithm implemented with a two-loop update
scheme~\cite{Ng11} for system size $N_{\rm s}=24\times24$ and
temperature $T=0.001$, while the estimated first-order transition
points somewhat deviate from the QMC values. This shows that the
present ED analysis can indeed give correct phase boundaries.
\end{thebibliography}
\end{document}